\documentclass[iop]{emulateapj}
\shorttitle{Spiral Density Waves in M81}
\shortauthors{Wang et al.}
\usepackage{bm}
\usepackage{sansmath}
\usepackage{mathrsfs}
\usepackage{epstopdf}
\usepackage[breaklinks,colorlinks,urlcolor=blue,citecolor=blue,linkcolor=blue]{hyperref}
\newcommand{\vect}[1]{\boldsymbol{#1}}
\newcommand{\hhwang}[1]{{\color{black} #1}}
\defcitealias{Fen14}{Paper I}

\begin{document}
%\linenumbers

\title{Spiral Density Waves in M81. II. Hydrodynamic Simulations for Gas Response to Stellar Spiral Density Waves}

\author{Hsiang-Hsu Wang\altaffilmark{1}, Wing-Kit Lee\altaffilmark{1}, Ronald E. Taam\altaffilmark{1,2}, Chien-Chang Feng\altaffilmark{1} and Lien-Hsuan Lin\altaffilmark{1}}
\email{hhwang@asiaa.sinica.edu.tw}
\altaffiltext{1}{Institute of Astronomy and Astrophysics, Academia Sinica, P.O. Box 23-141, Taipei 10617, Taiwan, R.O.C.}
\altaffiltext{2}{Department of Physics and Astronomy, Northwestern University, 2131 Tech Drive, Evanston, IL 60208, USA}

%\affil{Institute of Astronomy and Astrophysics, Academia Sinica, P.O. Box 23-141, Taipei 10617, Taiwan, R.O.C.}
%\email{hhwang@asiaa.sinica.edu.tw}

\begin{abstract}
Gas  response to the underlying stellar spirals is explored for M81 using unmagnetized hydrodynamic simulations. Constrained within the uncertainty of observations, 18 simulations are carried out to study the effects of selfgravity and to cover the parameter space comprising three different sound speeds and three different arm strengths. The results are confronted with those data observed at wavelengths of 8~$\mu$m and 21~cm.  In the outer disk, the ring-like structure observed in 8~$\mu$m image is consistent with the response  of cold neutral medium with an effective sound speed 7~km~s$^{-1}$, while for the inner disk, the presence of spiral shocks can be understood as a result of 4:1 resonances associated with the warm neutral medium with an effective sound speed 19~km~s$^{-1}$. Simulations with single effective sound speed alone cannot simultaneously explain the structures in the outer and inner disks. This justifies the coexistence of cold and warm neutral media in M81. The anomalously high streaming motions observed in the northeast arm and the outward shifted turning points in the iso-velocity contours seen along the southwest arm are interpreted as signatures of interactions with companion galaxies. The level of simulated streaming motions narrows down the uncertainty of observed arm strength toward larger amplitudes.
\end{abstract}


\section{Introduction}
The structures observed through gas and dust emissions in disk galaxies may shed light on the underlying processes associated with the formation of stars and the evolution of stellar disks.  Driven by the fact that grand design spiral arms are delineated with bright, young stellar associations, \citet{Rob69} suggested that star formation may be triggered at galactic scale as molecular clouds become gravitationally unstable when passing through galactic shocks.  The  two-armed spiral shock (TASS) solution pursued by \citet{Rob69} was based upon 
the assumption that the stellar arms appear as a normal mode which is presumably a quasi-stationary spiral structure (QSSS) \citep{Lin64,Lin66}. That is, the grand design spirals revolve about the galactic center with a well defined pattern speed and evolve slowly over several orbital times of the pattern. The gas response to the spiral density waves (SDWs) behaves in a predictable way, allowing direct comparisons between observations and predictions.

Numerical studies on the origin of self-excited spiral structures have recently been 
carried out by several authors \citep{Sel12,Don13} and reviewed by \citet{Sel14}. \citet{Sel14} argue that the spirals are growing cavity modes between resonances, at which the wave-particle interactions introduce abrupt changes to the impedance experienced by the traveling waves. The presence of resonances to the traveling waves is crucial for this mechanism to work. For M81, it appears that there is no inner Lindblad resonance associated with the observed pattern speed \citep{Adl96}. Thus, we do not take this view as the origin of spirals in M81. \citet{Don13} describe that apparent long-lived spiral as a result of the nonlinear development of gravitational wakes, which are initially seeded by density clumps, such as giant molecular clouds. In this view, the grand design spirals are nothing more than connections between self-perpetuating wave segments so that the visually `long-lived' structures are understood in a statistical sense. However,
persistent grand-design two-arm spiral structure are very difficult to produce numerically in an isolated 
disk \citep{Sel11}. Therefore, we believe that the QSSS hypothesis based on the theory of SDW remains relevant to this 
work. 

Several observations of disk galaxies with grand design spirals appear to support the presence of SDWs. The kinematic signatures in HI and CO qualitatively agree with the presence of galactic shocks \citep{Vis80b,She07},  the stellar age gradient across spiral arms is consistent with the picture of shock induced OB star formation \citep{Mar09,Pue97}, the angular offsets between dust and H$\alpha$ arms \citep{Ich85, Lo87,Til88,Nak94,Tos02,Egu04,Mou06} or dust and HI arms \citep{Tam08} are also expected from the theory of SDWs.

Aside from the aforementioned observed phenomena that are associated with the presence of major galactic shocks discussed by \citet{Rob69}, in comparison with the old stellar population in disks, the nonlinear 
gas response may lead to extra arms or substructures seen in optical images, which greater emphasize the young 
population I stars as well as ionized gas \citep{Blo91,Blo94,Elm99}. \citet{Elm80} suggested that the stellar spurs may be long-lived and share a common origin as the gaseous feathers. Here, the stellar spurs are understood 
as short stellar segments, which jut out from the main arms with large pitch angles, while the gaseous feathers are extinction features protruding downstream from the major shocks with pitch angles  similar to that of associated stellar spurs. Since the formations of young stars as well as giant molecular clouds are closely related to the gasdynamics in disks \citep{Elm14},  studying the nature of substructures in gas and dust is crucial to the understanding of disk evolution. 

Theoretical investigations on the origin of substructures primarily follow Roberts' picture within the framework of QSSS. Branches are identified as the secondary compression corresponding to 4:1 resonance \citep{Shu73}.  Feathers are considered to be a quasi-regularly spaced feature triggered by the gravitational instability associated with the underlying TASS solution \citep{Lee14,Lee12,Bal88,Kim02}.  More recently, normal mode analysis also suggests that the feathering phenomenon can be of a purely hydrodynamical origin due to the accumulation of potential vorticity which is generated in deformed shocks \citep{Kim14}.

QSSS is also the central working hypothesis that underlies some numerical works. Nonlinear gas response to an imposed spiral potential is followed numerically to take into account the effects of hydrodynamics, selfgravity, magnetic fields and multiphase medium.  Depending on the physics involved in the settings, feathers and spurs seen in numerical simulations are associated with the Magneto-Jeans Instability \citep[MJI,][]{Kim02,Kim06,She06}, the shearing-type instability \citep{Wad04}, overlapping of ultraharmonics \citep{Cha03}, and the sheared cold dense molecular clouds in the interarm region \citep{Dob06,Dob08}. It is, however, unclear whether one or more mechanisms are working under the conditions observed for real galaxies.  Numerical models built based on observational data may be the best way of distinguishing between different possibilities. 

Confrontation between theoretical predictions and the observations was recently carried out for the stellar disk of M81 by \citet[][hereafter Paper I]{Fen14}. Based on the modal analysis, two unstable modes with nearly equal growth rates were found for the two-armed spiral structure. The apparently separated inner and outer stellar arms \citep{Ken08} are identified to be consistent with the presence of the unstable mode that rotates about the galactic center at an angular speed of 25.5~km~s$^{-1}$~kpc$^{-1}$.   The predicted spiral shape, pattern speed as well as the trend of relative amplitude of arms are in good agreement with those inferred from observations. 

Complementary to the \citetalias{Fen14}, this second paper explores the gas response to the underlying stellar spirals using two-dimensional, unmagnetized, global hydrodynamic simulations. For the given mass model for M81, \citetalias{Fen14} provides the information about the spiral pattern, the pattern speed and the radial modulation of arm strength that are compatible to the reaction of the stellar disk. Under the assumption of QSSS, the two-armed stellar spiral pattern is imposed to rotate rigidly about the galactic center and the strength of the arms is scaled to be within the range of observational uncertainty. Since the total gas mass is only a small fraction ($< 1\%$, see Section 2.2) of the stellar mass, we ignore the 
back reaction by the gas on the stellar disk as a first approximation. Simulations with or without the selfgravity of gas are both considered because in the region of spiral shocks the selfgravity may be important to the stability of shocks \citep{Bal88,Lee12,Lee14}, the formation of branches \citep{Cha03} and the formation of self-bounded objects \citep{Dob08}. In the same spirit of \citetalias{Fen14}, the mass distribution of the gaseous disk as well as different effective temperatures that reflect the turbulent nature of gas are modelled based on observations \citep{Ian12}.

Compared to the earlier studies on the gas response in M81 \citep{Vis80a,Vis80b}, several aspects have been improved in this work. As reported in the \citetalias{Fen14}, our mass model for M81, spiral pattern, pattern speed and the modulation of arm strength are constrained by modern data and are understood within the framework of the theory of spiral density waves.  The properties of gaseous disks, such as the mass distribution and effective temperatures, are also modelled based on recent observations \citep{Wal08, Ian12}. Given all these constraints, the unmagnetized gas response  to the stellar spirals is followed for ten orbital times of the spiral pattern using hydrodynamic simulations with or without the selfgravity of gas.  Numerical results are confronted with the observed gas/dust spiral structures and HI kinematics. Substructures associated with the 4:1 resonances, ring-like structures, wiggle instability as well as the nature of the inner spiral shocks are explored and discussed. 

This paper is structured as follows. The physical models and a brief introduction of the numerical method are described in Section 2. Results obtained from the numerical simulations are presented and analyzed in Section 3. We discuss and conclude in Section 4. 

\section{Method of Analysis} 
\label{sec:method_analysis}
M81 (NGC~3031) is the major member of the nearby M81 galaxy group situated in the direction of Ursa Major and located at a distance 3.6~Mpc \citep{Ken08,Kar05}.  The disk of M81 is tilted with an inclination angle of roughly $60^{\circ}$ in the sky, with inclination $0^{\circ}$ corresponds to a face-on disk.  In this work, we adopt an inclination $58^{\circ}$ \citep{Adl96, deV91} to tilt our face-on results when comparing with observations. The proximity and the favorable inclination make M81 one of the well studied spiral galaxies for both disk structure and gas kinematics. In optical images, a pair of nearly bi-symmetric grand design spirals is prominent in the disk. The spiral tail ending in the north will be called northeast arm and the one ending in the south is called southwest arm. 

We perform two-dimensional, hydrodynamical simulations with or without selfgravity of gas to investigate the nonlinear gas response to the grand-design stellar spirals in M81. In \citetalias{Fen14}, a bulge-disk-halo mass model is constructed to support the observed rotation curve. For the given mass model, two spiral modes with nearly equal growth rates were found for two-armed spirals using modal analysis. The mode $(m=2,n=1)$, where $m$ denotes the number of arms and $n$ is the number of nodes, were identified to have the appropriate characteristics that fit the observed stellar arms. This particular spiral mode was presumably `selected' due to the recent interactions with M82 and NGC 3077 as hinted by the tidal tails observed in neutral gas \citep{Tho93,Yun94,Yun99}. Superimposed on the stellar disk built for M81, the pattern of spiral perturbation, the associated angular speed and arm strength adopted in our simulations are constrained by observations and are derived naturally from the modal analysis. The good agreement between analytic predictions, though still subject to uncertainty in the radial profile of stellar velocity dispersion, and observations gives the confidence that the theory of spiral density wave is one of the plausible mechanisms that may explain the origin of spirals in M81.

Although the gas behavior and the stellar dynamics may couple to each other through the combined Poisson equation, the following two differences between gas and stellar disks enable us to consider them separately for this specific galaxy, M81. First, the total mass in gas is less than 1\% of the stellar mass. The gravity of gas has negligible influence on the stellar dynamics. It is therefore adequate to treat the gas as a passive component reacting to a potential field dominated by the stars and the dark matter halo. Nevertheless, the bright, young stars observed along galactic shocks indicate that the selfgravity of gas in dense regions should play a role in forming stars. Second, the fact that azimuthal cuts of the broad, smooth stellar arms can be well fitted with a single sinusoidal function \citep{Ken08} justifies the linear analysis for the stellar disk. On the other hand, the presence of opaque dust lanes along the galactic shocks reflects the dissipative nature and nonlinear behaviors of the gas. The linear analysis for the stellar dynamics of M81 has been reported in the \citetalias{Fen14}, while in this paper we focus on the nonlinear gas response using the hydrodynamic code, Antares. 

\subsection{Numerical Method}
The numerical simulations are carried out with a high-order Godunov code known as Antares, in which the hydrodynamic fluxes on zone interfaces are calculated with the exact Riemann solution \citep{Yua05}. Cartesian coordinates are adopted to avoid the need to impose an inner boundary condition to handle the coordinate singularity at the origin. The origin of the coordinates coincide with the galactic center. At the outer boundary, a radiation boundary condition is imposed. That is, the wave characteristic decomposition is performed at each boundary cell, allowing waves to propagate outwards, but suppressing incoming waves. With this treatment, reflection is automatically prohibited at the boundary. 

The gas disk is evolved in a frame corotating with the stellar spirals and the turbulent nature of gas is approximated using an isothermal equation of state. The gas response is solved numerically with the following hydrodynamical equations:
\begin{eqnarray}
   \frac{\partial \sigma}{\partial t}+\nabla\cdot(\sigma\vect{v})=0, \\
   \frac{\partial \vect{v}}{\partial t}+\vect{v}\cdot\nabla{\vect{v}}=-\frac{\nabla P}{\sigma}-\nabla V+\Omega_p^2 r-2\Omega_p \times \vect{v},
\end{eqnarray}
where $\sigma$ is the surface density of gas, $\vect{v}$ denotes the velocity vector, $P$ is the vertically integrated pressure, $r$ represents the galactocentric distance measured in the plane of galaxy, $\Omega_p$ is the pattern speed and $V$ the total gravitational potential composed of three components:
\begin{equation}
   V = V_0+V_1+V_g. 
\end{equation}
$V_0$ denotes the axisymmetric component (disk-bulge-halo) that supports the rotation curve; $V_1$ represents the perturbation potential attributed to the presence of stellar spirals and $V_g$ is the component associated with the selfgravity of the gas. To avoid the noise due to the sudden introduction of the stellar spirals, the arm strength is turned on slowly, reaching full strength at one orbital time of the pattern and is kept constant thereafter. In order to close the equations, the pressure, $P$ is related to the surface density of gas through $P=c_{\rm s}^2 \sigma$, with $c_s$ corresponding to the effective isothermal sound speed that reflects the turbulent nature of gas. For a given mass distribution of imposed perturbation, the forcing $-\nabla V_1$ can be calculated using the Poisson solver described below. Since the perturbation is static in the corotating frame, the forcing associated with the stellar spirals needs to be calculated only once at the beginning of the simulations. The gravitational potential of the infinitesimally thin gas disk is included through the relation:
\begin{equation} \label{eq:poisson_eq}
 \nabla^2 V_g=4\pi G \sigma \delta(z),
\end{equation}
where $G$ is the gravitational constant and $\delta$ is the Dirac delta function. 

Solving the Poisson equation, i.e., Equation~(\ref{eq:poisson_eq}), for an infinitesimally thin gaseous disk is intrinsically a three-dimensional problem. However, a direct sum calculation for the potential function, $V_g$, is avoidable. Specifically, the force field in the plane of the disk is calculated by integration, where the force is written as a double summation of the product of surface density and a kernel, which is in a form of convolution. The computational complexity associated with the convolution of length $N$ is reduced from $O(N^2)$ to $O(N\ln N)$ using the fast Fourier transform (FFT). As a result, the entire force computation of our two-dimensional problem has a complexity of $O((N\ln N)^2)$. We note that the FFT is only used to accelerate the computation, and our Poisson solver does not require a periodic boundary condition. We refer to the work by \citet{Yen12} for those who are interested in this computational technique. 

\subsection{Models and Parameters} 
\begin{figure}
\epsscale{1.1}
\plotone{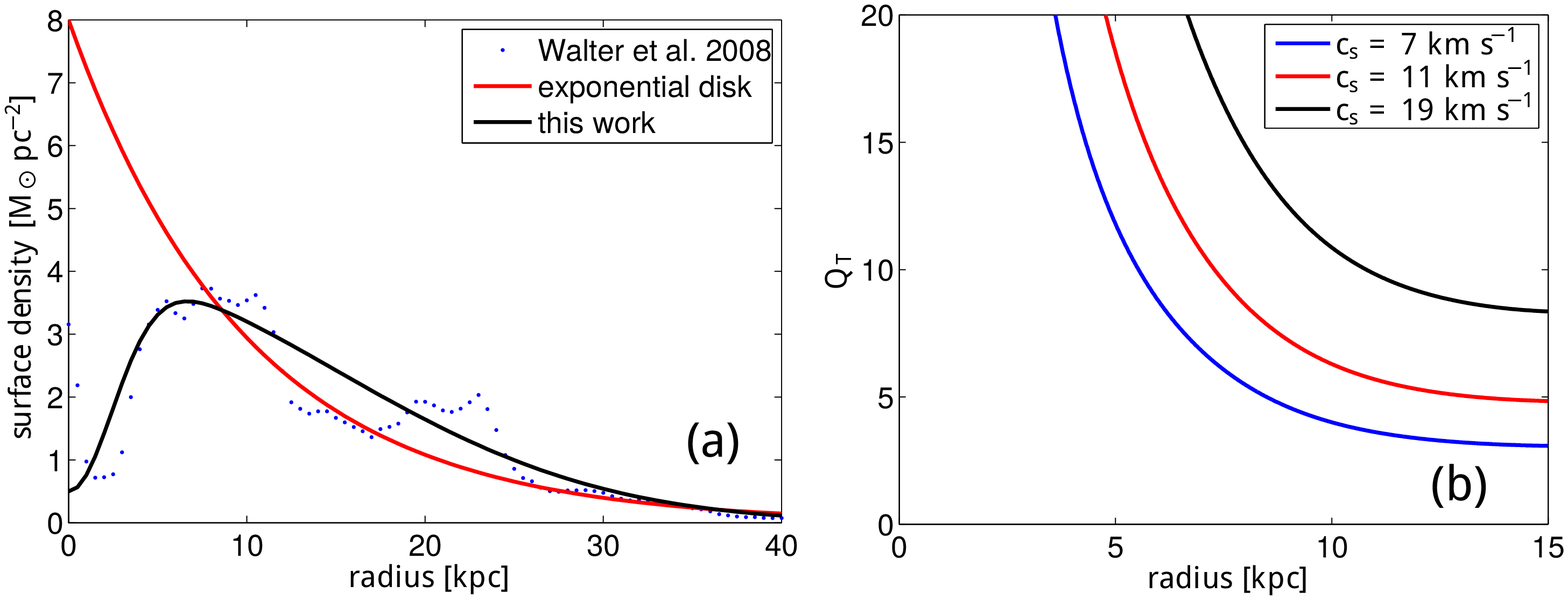}
\caption{(a) Azimuthally averaged gas surface density as a function of galactocentric distance. The blue dotted line is derived from the $0$th-moment HI map \citep{Wal08} averaged over circular rings 500~pc wide in the plane of M81 and multiplied by a factor 1.4 for heavier element content. The red line is a fitting to the observed data using an exponential function of a scale length 10~kpc, while the black line represents the initial gas surface density adopted in this work described by Equation~(\ref{eq:gas_distribution}). (b) Toomre $Q$ as a function of radius for models with sound speeds $c_s=7,~11,~19$~km~s$^{-1}$, respectively. \label{fig:surden_Q}}
\end{figure}

The evolution of gas is calculated within a rectangular domain with a physical length 24~kpc on each side. The calculation domain is uniformly subdivided into $1024 \times 1024$ cells, corresponding to a spatial resolution $\approx 23$~pc or 1.32\arcsec~at a distance of 3.6~Mpc. The simulations are initialized with an axisymmetric gas disk rotating about the galactic center. Figure~\ref{fig:surden_Q}a shows the observed gas distribution as the dotted curve, which is derived from the 0th-moment HI map \citep{Wal08} averaged over circular rings 500~pc wide in the plane of the galaxy and multiplied by a factor of 1.4 for helium and heavier element content \citep{Cro07}. Except for the very central region ($r<1$~kpc), the inner 4~kpc zone is HI depleted \citep{Adl96,Rot75b}. The outer region ($r>9$~kpc) can be well fitted with an exponential function (red curve) with a scale length of 10~kpc \citep{Yun94}. The initial gas distribution adopted in this work is shown as the black curve and is described by:
 \begin{equation} \label{eq:gas_distribution}
    \sigma(r) = 4\exp(-\frac{r^2}{2a_1^2})-3.5\exp(-\frac{r^2}{2a_2^2}),
 \end{equation}
with $\sigma$ being the surface density of the gas disk in units of $M_{\odot}$~pc$^{-2}$, $a_1=15.0$ and $a_2=2.5$~kpc. The total mass enclosed within the radius of outer 4:1 resonance, i.e., 10.8~kpc, is estimated to be $1.15\times 10^{9}$~$M_\odot$. 

The interstellar neutral gas is multiphase and turbulent. Both thermal and kinematic broadening contribute to the observed absorption or emisson line widths. While the cold neutral medium ($\approx 100 K$, CNM) associated with narrow line profiles appears to be clumpy and well correlated with star forming regions, the warm neutral medium (a few thousand $K$, WNM) tends to be smoothly and ubiquitously distributed \citep{You96,You97,You03}. \citet{Ian12} analyzed the HI super velocity profiles of nearby galaxies using a stacking method that is usually applied to construct high signal-to-noise profiles. They found that these super profiles are best described using a narrow and a broad Gaussian component, consistent with the presence of CNM and WNM. For M81, the velocity dispersion is estimated to be $6.7\pm 0.1$~km~s$^{-1}$ for the CNM and $19 \pm 0.2$~km~s$^{-1}$ for the WNM. The velocity dispersion derived from the one-component Gaussian fit gives $11.2\pm 0.2$~km~s$^{-1}$. In the cases of low surface density, i.e., the selfgravity of gas is unimportant, \citet{Dob08} has shown that responses of CNM and WNM to imposed spiral arms are dynamically decoupled. The major effect of WNM is to confine the substructures, which would anyway happen in the cold-component only simulations, to higher density. Since the surface density of gas in M81 is not particularly high, we study the responses of cold and warm gases separately.  As a comparison, simulations using the effective sound speed that corresponds to one-component Gaussian fit are also performed. In these cases, we implicitly assume that cold and warm gases are dynamically well coupled.

The initial rotation velocity of gas follows the rotation curve modeled in \citetalias{Fen14} (see Figure~2 therein), where a three-component mass distribution associated with a stellar disk, a dark matter halo and a stellar bulge was constructed based on observational data. Given the gaseous surface density, sound speeds and the rotation curve, the stability of a gaseous disk is described by the Toomre $Q_{\rm T}$ defined as:
\begin{equation}
   Q_{\rm T} = \frac{c_s \kappa}{\pi G \sigma},
\end{equation}
where $\kappa$ is the epicyclic frequency associated with the rotation curve. As shown in Figure~\ref{fig:surden_Q}b, all the gaseous disks are initialized with $Q_{\rm T}$ well above unity, i.e., the initial states of gas disks are expected to be stable to axisymmetric perturbations.

Superimposed on the axisymmetric stellar disk is a stellar spiral pattern rigidly rotating about the galactic center with an angular speed of $\Omega_p=25.5$~km~s$^{-1}$~kpc$^{-1}$. The spiral pattern and the associated pattern speed are calculated self-consistently with the three-component model based on the mode analysis \citepalias{Fen14}. The apparently separated inner and outer spirals observed by \citet{Ken08} are identified as an unstable spiral mode, thereby rotating with the same angular speed. The radii of resonances depends on the effective sound speeds and will be discussed in Section~3.2. Here, we note that no inner Lindblad resonance, i.e., 2:1 resonance, is associated with the pattern speed adopted in this work. 

\begin{figure}
\epsscale{0.5}
\plotone{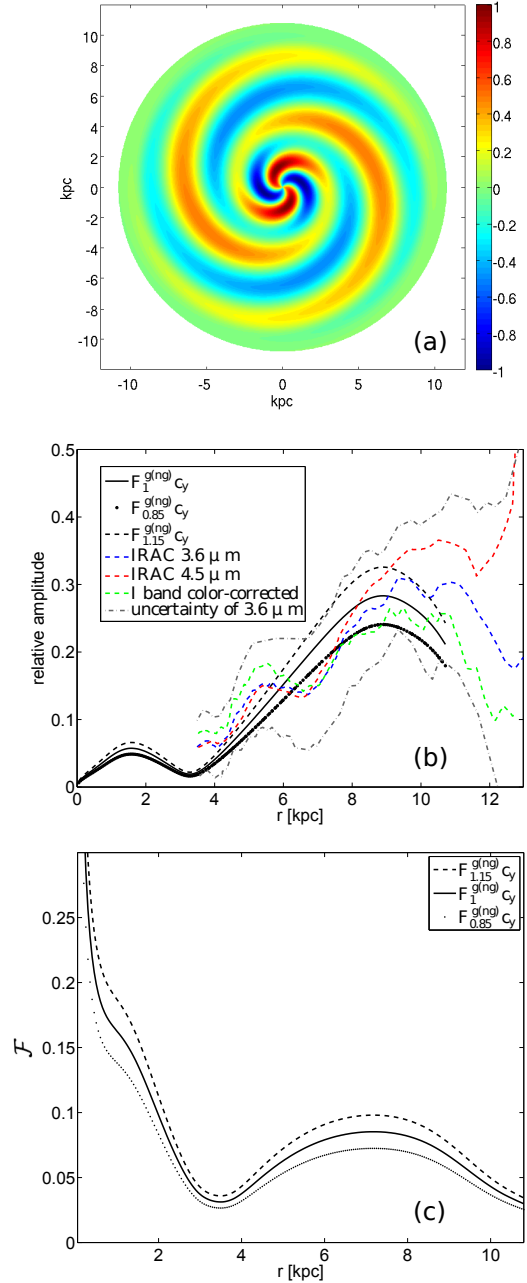}
\caption{ \label{fig:spiral_amplitude} (a) Density distribution of stellar spirals superimposed on an axisymmetric stellar disk described in the \citetalias{Fen14}. The density shown in this image is normalized and only the contrast is emphasized. Positive values (toward the red end) represent density excess while negative values (toward the blue end) are density deficit with respect to the azimuthal average. (b) Relative amplitude of spiral density wave as a function of radius. \hhwang{For a given radius, the relative amplitude is defined as the ratio of the amplitude of azimuthal variation to the azimuthal average. This definition reflects the observed mass variation in stellar components.} The grey dash-dotted curves bracket the uncertainty of the data at 3.6~$\mu$m. The spiral pattern and the modulation of arm strength are taken from the modal analysis for M81 and are scaled to fit the observed arm strength. The solid black curve is the same as the one reported in \citetalias{Fen14} and is taken as the fiducial arm strength $F^{g(ng)}_1c_y$. The model that corresponds to 15\% increase (decrease) of the fiducial model is shown in dashed (filled-circle) black line and is labeled $F^{g(ng)}_{1.15}c_y$ ($F^{g(ng)}_{0.85}c_y$). \hhwang{(c) The radial profiles of arm strength, $\mathcal{F}(r)$, defined as the ratio of maximum arm force to the centrifugal force.}}
\end{figure}

The mass distribution of the imposed stellar spirals is shown in Figure~\ref{fig:spiral_amplitude}a, where the surface density is normalized and only the density contrast is emphasized. The strengths of stellar spirals are constrained with observations as shown in Figure~\ref{fig:spiral_amplitude}b. Within the uncertainty of the observed $3.6\mu$m data, gas reactions to different strengths of density waves are explored. The black solid is taken as the fiducial strength of spirals, while the model corresponding to a 15\% increase (decrease) in strength with respect to the fiducial one is shown by dashed (dotted) black curve. In this plot, the curves of models are shown to the radius of outer 4:1 resonance, where the outgoing boundary condition is applied for the mode analysis. Beyond this particular radius, the strength of spirals is assumed to exponentially decay to nearly zero at 12~kpc to mimic the absorption of short trailing waves at the outer Lindblad resonance. 

\hhwang{Figure~\ref{fig:spiral_amplitude}c shows the radial profiles of arm strength, $\mathcal{F}\equiv f_{\rm sp, max}(r)/r\Omega^2(r)$, in terms of the ratio of the maximum arm force to the centrifugal force \citep{Shu73}. We note that the Equation~(4) of \citet{Shu73} is a useful dimensionless measure for arm strength when spirals are tightly wound, so that the nonlinear gas responses can be followed with the small pitch angle approximation. The outer spirals of our model are tightly wound with pitch angle no more than 18$^{\circ}$. The arm strengths of the outer spirals are peaked at $7.2$~kpc with a ratio of 7.2\%, 8.5\% and 9.8\%, respectively, for the dotted, solid and dashed curves. On the other hand, the pitch angle of inner spirals grows rapidly with decreasing radius and the small pitch angle approximation is no longer valid for the analysis by \citet{Shu73}. When approaching the galactic center, the term $1/r$ appearing in the definition of $\mathcal{F}$ leads to the rapid growth as shown in Figure~\ref{fig:spiral_amplitude}c.}

In this work, the gas response to the underlying stellar spirals are explored using three different sound speeds and three different spiral strengths. Including the models with or without selfgravity, 18 simulations are performed and analyzed. Different models are denoted with a symbol like $F_{x}^{g(ng)}c_y$, where $x=1.15,~1,~0.85$ corresponds to the three strengths of stellar spirals, $y=7,~11,~19$ stands for the sound speeds of gas and the superscript is used to distinguish between selfgravitating ($g$) and non-selfgravitating ($ng$) models. Among them, the self-gravitating simulation with $(x,y)=(1,11)$, i.e., $F^g_{1}c_{11}$ , is taken as the fiducial model.  

\section{Results} 
Gas response to the spiral density waves is explored using a parameter space constrained within the uncertainty of observations. Observationally, emission or absorption from different tracers are used to probe different physical conditions and environments in galaxies. Dust emissions at infrared wavelengths are thought to trace the dense regions associated with star formation activity or the presence of galactic shocks. \citet{LaV06} have shown that emission at 8~$\mu$m suffers less extinction compared to optical wavelengths and is therefore an ideal tracer for galactic shocks and feathers which may be well extended into the interarm region. However, dust continuum is not ideal for deriving kinematics and furthermore M81 seems also to lack strong molecular emissions over the disk \citep[][ and references therein]{San11}. Fortunately, \citet{San11} compared the spectral profiles of CO ($J=3-2$) with those of HI data in M81 and found good agreement between peak velocities and line widths in both cases, indicating that the kinematic conditions of molecular and atomic gas are similar. We note that those strong CO ($J=3-2$) emissions they detected are clustered only in two small regions within the spiral arms. The neutral atomic gas (HI), on the other hand, is more widespread and abundant over the disk of M81, except for the central region. The hydrogen 21-cm line is therefore suitable for mapping detailed gas kinematics over the disk. 

\subsection{HI-to-Dust Conversion Function}
\begin{figure}
\epsscale{1}
\plotone{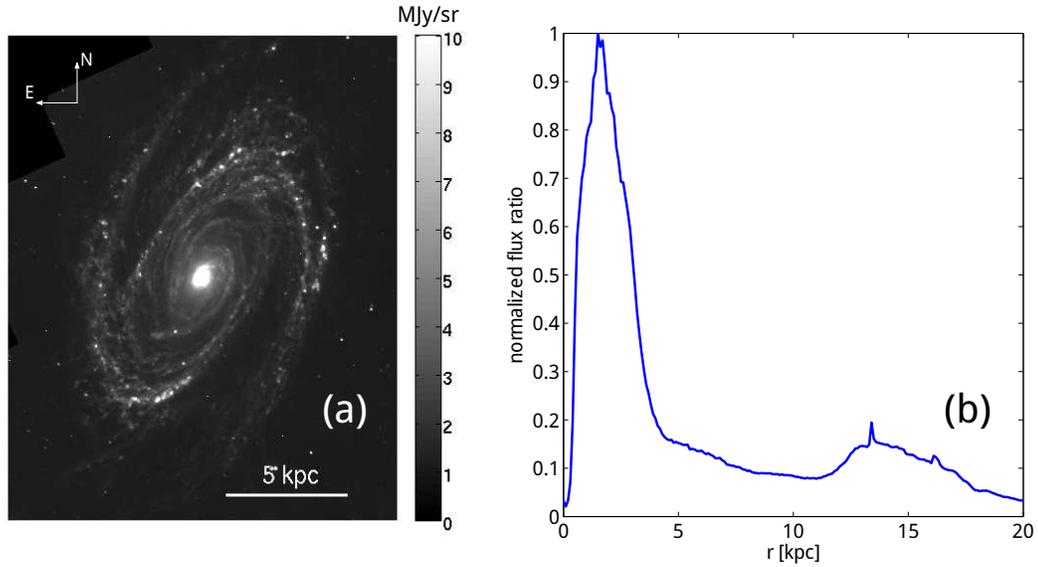}
\caption{ The image of dust emission observed with the IRAC {\it Spitzer} at the wavelength of 8$\mu$m \citep{Wil04}. (b) Normalized flux ratio between observed emissions of dust and HI gas as a function of radius. \label{fig:8um_observation}}
\end{figure}

Figure~\ref{fig:8um_observation}a shows the image of M81 at the wavelength 8~$\mu$m observed with the Infrared Array Camera (IRAC) on board the {\it Spitzer Space Telescope} \citep{Wil04}. The overall structures include two pairs of grand-design spirals, which can be roughly separated with the radius $r=4$~kpc. For the convenience of discussion hereafter, the region inside (outside) 4~kpc is referred as the inner (outer) disk and the corresponding spirals are called the inner (outer) spirals. 

It has been shown that infrared emission is strongly correlated with the optical dust lanes, which presumably track galactic shocks. Visually, the outer spirals appear to be clumpy and filamentary, and are roughly outlined by those bright dots shown in Figure~\ref{fig:8um_observation}a. The prominent feathering substructures are generally found downstream of the trailing spirals in the outer disk within corotation. In contrast to the outer spirals, the inner spirals, which reside in the region devoid of HI (see Figure~\ref{fig:HI_vLoS}a), are relatively smooth and well organized. While the gas may be highly compressed in both the inner and outer spiral shocks, the outer spirals may be experiencing some instabilities associated with the shocks and therefore facilitate the formation of dense clumps as well as the feathers. 

Since our simulations evolve the response of total gas, a function that describes the dust-to-gas ratio is required for the comparisons between numerical results and the observed structures of dust. We extract the conversion factor by comparing the 0th-moment maps at the wavelengths of 8~$\mu$m \citep{Wil04} and 21~cm \citep{Wal08}. The conversion factor is a function of radius and is defined to be the total intensity ratio, $f(r)=I_{\rm dust}(r)/I_{\rm HI}(r)$, for a given circular ring of 200~pc wide centered at galactocentric radius $r$ in the plane of galactic disk. The resultant $f(r)$ shown in Figure~\ref{fig:8um_observation}b is normalized with the peak value \hhwang{and therefore dimensionless}. Consistent with the impression from the images, $f(r)$ indicates that the inner disk is relatively dust rich compared to the outer disk. \hhwang{We note that when applying the dimensionless $f(r)$ to construct the 8~$\mu$m images based on the simulated responses of the total gas, we are concerned only with the radial distribution of the dust so that the structures in the dust component can be properly emphasized. }

\subsection{4:1 Resonances}
The radius that corresponds to ultraharmonic resonance is a function of sound speed and can be found when the following relation is fulfilled \citep{Shu73}:
\begin{equation}
  \nu^2-\eta=\frac{1}{n^2},
\end{equation}
where $\nu \equiv m(\Omega_p-\Omega)/\kappa$ is the dimensionless Doppler shifted frequency, $m$ denotes the number of spiral arms, $\eta=m^2c_s^2/r^2\kappa^2\sin^2(i)$, $i$ stands for the pitch angle of spiral arms and $n$ is an integer corresponding to the mode of harmonics. For instance, $n=1$ corresponds to Lindblad resonances and $n=2$ to 4:1 resonances. Negative $\nu$ marks those resonances inside corotation, while positive $\nu$ for those beyond corotation. We note that no inner Lindblad resonance (ILR) is associated with the pattern speed used in our models. 

\begin{figure}
\epsscale{1}
\plotone{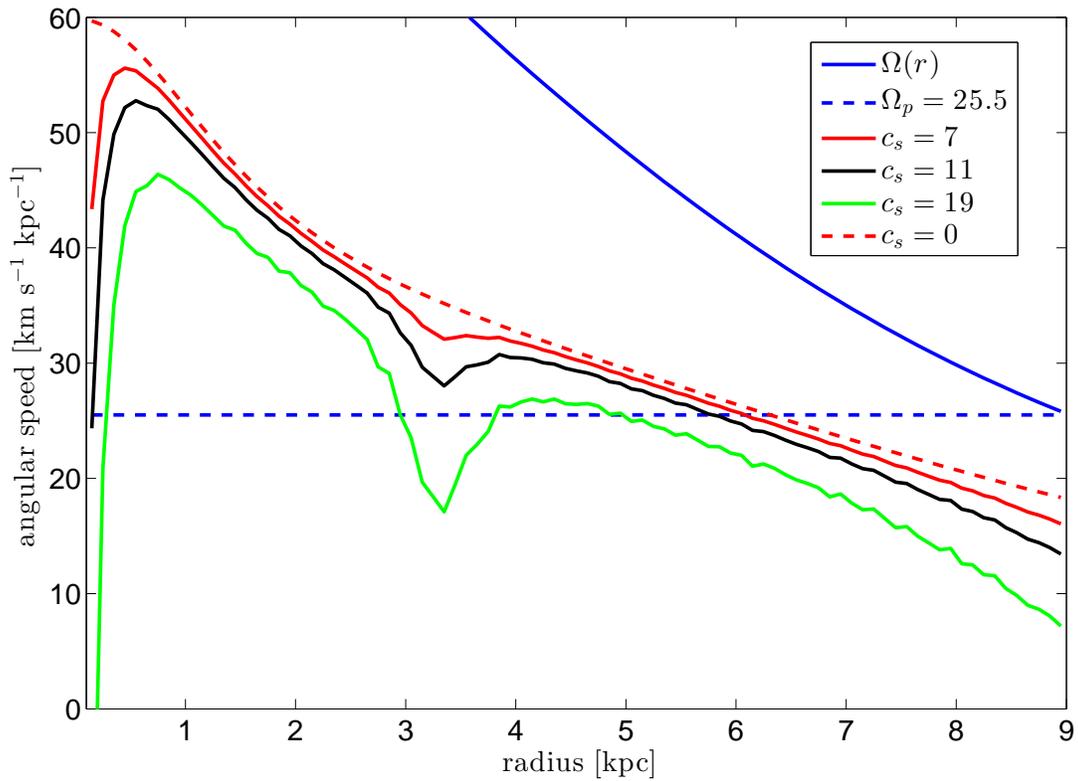}
\caption{The locations of 4:1 resonances for models with $c_s=0$, 7, 11, 19~km~s$^{-1}$. The solid blue curve represents the angular speed as a function of radius and the dashed blue stands for the pattern speed. See the text for the function form for the curves of different sound speeds. The intersections with dashed blue line mark the locations of 4:1 resonance. They are 6.3, 6.1, 5.8 and (4.9, 3.8, 3.0)~kpc for $c_s=0$, 7, 11, 19~km~s$^{-1}$, respectively. \label{fig:ultraharmonics}}
\end{figure}

The exact locations of 4:1 resonance for different sound speeds used in our models can be read from Figure~\ref{fig:ultraharmonics}. The solid-blue curve is the angular speed, $\Omega(r)$, as a function of galactocentric distance. The dash-blue line represents the pattern speed $\Omega_p=25.5$~km~s$^{-1}$~kpc$^{-1}$. The dashed red, solid red,  black and green curves are described by $\Omega-(\kappa/2)\sqrt{1/4+\eta}$ and corresponding to sound speeds 0, 7, 11, 19~km~s$^{-1}$. Their intersections with the dash-blue line mark the locations of 4:1 resonances. They are 6.3, 6.1, 5.8 and (4.9, 3.8, 3.0)~kpc for $c_s=0$, 7, 11, 19~km~s$^{-1}$, respectively. In particular, multiple 4:1 resonances are found for models with $c_s=19$~km~s$^{-1}$.  Here, we note that the curve associated with the pressure free case, i.e., $c_s=0$,  is presented as a reference to show the degree of shift due to the pressure effect. The very inner intersections are ignored since we do not have enough spatial resolution to resolve those resonances located near the center, though they might be also scientifically interesting. The dips seen in the curves are due to the drop of pitch angle between 3 and 4~kpc. It is evident that the locations of inner 4:1 resonances are shifted inward in the models with higher sound speeds. \hhwang{The inward shifted 4:1 resonances with increasing sound speed can be physically understood in terms of an increasing effective epicyclic frequency. In addition to the purely kinematic epicyclic motion, which is a result of the conservation of angular momentum, the effective pressure provides additional restoring force that tends to push material back toward its guiding center. As a consequence, those curves associated with non-zero sound speeds are shifted downward as shown in Figure~\ref{fig:ultraharmonics} and the corresponding locations of resonances are moved inward. }
\subsection{The Outer Disk}
\subsubsection{Comparisons with the 8~$\mu$m image}
On the left of Figure~\ref{fig:8um_comparison} are the snapshots of simulated dust emission for the models as denoted on the upper-left corner of each image, which is presented by tilting the face-on image with an inclination angle of 58$^\circ$. Those images are taken at the temporal point corresponding to 5 revolution times of the spiral pattern after the conversion function $f(r)$ is applied. The gas response has reached quasi-steady state and does not change rapidly with time. By applying the conversion function, we implicitly assume that the kinematics of dust is well coupled to that of the total gas. The white-dashed ellipses mark the radii of inner 4:1 resonances for each model. The radii correspond to ultra-harmonics are shifted inward toward the galactic center for those models with higher sound speeds. In particular, the model $F^{g}_{1.15}c_{19}$ has three radii that correspond to 4:1 resonances. The right column of Figure~\ref{fig:8um_comparison} shows the direct comparisons by overlaying those images shown in the left column over the 8~$\mu$m image shown in Figure~\ref{fig:8um_observation}a.

As can be seen in the left column of Figure~\ref{fig:8um_comparison}, qualitatively, the overall structure of the outer disks are similar except for the following differences. The outer spirals stretch out more toward the radius of corotation in the models with lower sound speeds. Spiral shocks are expected to diminish when approaching the corotation because the imposed spiral perturbation becomes subsonic there. The extent of the subsonic region increases with sound speed. It is therefore not surprising that the outer spirals are less extended in those models with higher sound speeds. We note that this is not generally true if the imposed spirals are sufficiently strong that the perturbed velocity allows 
shocks to develop even at the corotation radius \citep{KimY14}.

Perhaps, the most distinct difference is the presence of substructures. Figure~\ref{fig:8um_comparison}a ($F^{g}_{1.15}c_7$) shows feathers emanating from the primary shocks, reminiscent of those observed filamentary (Figure~\ref{fig:8um_observation}a) substructures protruding from the outer spirals with high pitch angles. The spiral shocks fragment to form dense clumps with quasi-regular spacing at a scale of kpc. The clumps follow the radially inward streaming motions along the spiral shocks and eventually leave the arm region due to the conservation of angular momentum. Upon leaving the arm region, those clumps are torn apart in the form of elongated trailing feathers due to the galactic shear. In contrast to Figure~\ref{fig:8um_comparison}a, no feathering substructure is found for those models with higher sound speeds shown in Figures~\ref{fig:8um_comparison}b and \ref{fig:8um_comparison}c. The discussion on the possible origins and implications of feathers appearing in M81 is deferred to Section~4.1. 

The other kind of substructure, which is also prominent in the outer disks, is the secondary compression associated with inner 4:1 resonances. The secondary compression appears as branch-like spirals with pitch angle slightly smaller than the primary ones. In all cases, it is evident that the ultraharmonic waves emerge at the resonant radii (as indicated by the white ellipses) and propagate inwards. Figure~\ref{fig:8um_comparison}d suggests that the observed large ring-like structure (at $r=5$~kpc) as seen in the 8~$\mu$m image can be understood as the feature of 4:1 resonance. Although Figures~\ref{fig:8um_comparison}b and \ref{fig:8um_comparison}c also show the similar feature, those ring-like structures are shifted toward smaller radii (at $r=3.8$ and 0.9~kpc, respectively) and therefore do not match well with the ring observed in the outer disk. 

We do not intend to overlay all the snapshots obtained from 18 models over the observed image because they are visually similar for those models with the same sound speeds $c_s=11$ and 19 except for $c_s=7$ ~km~s$^{-1}$.  Figure~\ref{fig:image_array_c7} shows the face-on snapshots for models with $c_s=7$~km~s$^{-1}$ at $t \approx 1.2$~Gyr.  The upper panel are models that take into account the selfgravity of gas, while those shown in the lower panel do not. The strength of spiral forcing increases from the left to the right. Within the uncertainty of the observed arm strength, feathers start to emerge in the selfgravitating model with fiducial arm strength and become increasingly prominent upon increasing arm strength. Comparing to the non-selfgravitating cases, while feathers are also prominent in the model with the strongest arm strength, no feathers are found for the case with fiducial arm strength (lower middle). On the other hand, the density of clumps found in the selfgravitating cases is in general higher than those found in the non-selfgravitating cases. The differences and similarities between the upper and the lower panels suggest that selfgravity facilitates, but is not critical to the occurrence of feathers in the simulations. 

\subsubsection{Effects of sound speed, arm strength and selfgravity}
By examining the details of azimuthal cuts for a given radius, Figure~\ref{fig:density_cut_r5p5} explores the effects of sound speed, strength of spiral forcing and the gaseous selfgravity.  The left column shows the cuts of surface density and the right one shows the corresponding streaming motions for the radii ($r=5.5, 2.5$~kpc) as denoted in the plots at $t=450$~Myr for Figures~\ref{fig:density_cut_r5p5}a, \ref{fig:density_cut_r5p5}b, \ref{fig:density_cut_r5p5}c (nearly 2 orbital times of the spiral pattern) and at $t=1.2$~Gyr for Figure~\ref{fig:density_cut_r5p5}d (10 orbital times of the spiral pattern). At this particular temporal point, $t=450$~Myr, chosen just before the wiggle instability sets in, the gaseous spirals in all models are well developed and still smooth for analysis. The streaming motion (in the plane of galaxy) is defined to measure the magnitude of velocity departure from the initial axisymmetric velocity field. Since these azimuthal cuts are inside the corotation radius, the gases flow from the left (negative phase) toward the right (positive phase) of plots.  The zero phase corresponds to the density peak of the imposed stellar spirals at the given specific radius. The information is duplicated in a full $2\pi$ azimuthal range, thereby only the phase range [$-\pi/2$ $\pi/2$] is shown.  
 
From Figure~\ref{fig:density_cut_r5p5}a,  it is evident that gases of different sound speeds respond differently to the same strength of underlying spiral forcing. The density \hhwang{enhancement} in cold gas is higher than that of the warm gas by a factor of three. The density \hhwang{enhancement} is defined as the ratio of the peak surface density and the azimuthally averaged surface density for a given radius.  The second compression is prominent in the models $F^{g}_1c_{7}$ and $F^{g}_1c_{11}$, while absent in the model $F^{g}_1c_{19}$. This is because the radius $r=5.5$~kpc is inside and not far from the 4:1 resonances of the former two models, but outside the outermost 4:1 resonance of the latter case (see Figure~\ref{fig:ultraharmonics}). The overall structure is shifted downstream for colder gas. The phase offset of peak surface density between the cold ($c_s=7$~km~s$^{-1}$) and the warm ($c_s=19$~km~s$^{-1}$) gases is significant.  The offset in phase can be more than 20$^\circ$, corresponding to spatial separations more than 2~kpc for the radial range 4 to 6.5~kpc . No shock is found for the model $F^g_1c_{19}$ when $r>6.5$~kpc. The peak separation between cold and warm gases within the annulus ranging from $4<r<6.5$~kpc might be observable if the stacking method used to derive the single superprofile for one specific galaxy \citep{Ian12} can be applied to the annulus by stacking information radially with phase correction for the shock locations. In Figure~\ref{fig:density_cut_r5p5}b, for a given sound speed, a stronger spiral forcing tends to enhance the density contrast but the overall profiles are similar. Due to the low gaseous surface density in M81, selfgravity does not have significant role in shaping the density profile as shown in Figure~\ref{fig:density_cut_r5p5}c. 

The corresponding streaming motions for those models in the left column are shown on the right of Figure~\ref{fig:density_cut_r5p5}. For the outer disk, models with lower sound speeds or stronger forcing tend to have higher streaming velocities. The difference due to selfgravity is not significant. The spikes corresponding to the rapid changes in velocity fields trace the location of shocks. The second compression is also well correlated to the rapid changes in streaming motions. The maximum values of streaming motions measured for the outer disk in all models are no more than 31~km~$s^{-1}$. Specifically, the peak values of streaming motions for models with forcing larger than or equal to that of fiducial case lie somewhere between 25 to 30~km~s$^{-1}$, which are well consistent with the observed values 25 to 30~km~s$^{-1}$ in the plane of galaxy for the southwest arm but far below the values (up to 50~km~$^{-1}$) measured for the northeast arm \citep{Adl96}. The streaming motions obtained for the models with the weakest forcing are well below 25~km~s$^{-1}$. The comparison between streaming motions obtained from simulations and that measured for the southwest arm favors a forcing stronger than our fiducial case, narrowing down the range of uncertainty in the arm strength. On the other hand, simulation results indicate that the observed ultra-high streaming motions in the northeast arm is not consistent with the observed arm strength.  The result shown in Figure~\ref{fig:density_cut_r5p5}d is intriguing and is postponed to the discussion for the inner disk.  
 
\subsubsection{Comparisons with the HI kinematics}
In Figure~\ref{fig:HI_vLoS}a, the observed HI intensity map (colored in red) is overlapped on the 8~$\mu$m image  (colored in gray) shown in Figure~\ref{fig:8um_observation}a. The white ellipses represent the corotation radius ($r=9$~kpc) and $r=4$~kpc, inside which the 21~cm detection is low so that the iso-velocity contours shown in Figure~\ref{fig:HI_vLoS}b are noisy. Hereafter in this subsection, our discussions for kinematics will only focus on the outer disk, i.e., the region outside the radius of $4$~kpc.  Overall, the distributions of dust and neutral gas are correlated well inside the corotation radius. The dust, which presumably traces the galactic shock, in the southwest arm is well organized until $r=8$~kpc. The diminishing of shocks before the radius of corotation is expected and is consistent with the pattern speed adopted in this work. On the other hand, the HI in the southwest arm extends well beyond the corotation radius all the way to the bright blobs shown in the southeast. The southwest HI arm appears to be flocculent and widespread for the part outside corotation and has no counterpart of dust emission. Assuming that the motion of dust is well coupled to that of HI gas, the different distribution of dust and HI gas indicates that the source of HI, which lies beyond corotation, is likely external to rather than stripped from the inner disk. In other words, the southwest arm may have just experienced an interaction presumably with its companions in the recent past.

Inside the corotation radius, a branch-like substructure downstream the southwest arm is prominent in both dust and neutral gas. No counterpart of this substructure is found for the  northeast arm and our simulations do not reproduce this feature either. Unlike the above mentioned extended HI spiral outside the corotation radius, this branch-like substructure is prominent in both HI and dust emissions. We suspect that this substructure is a tidal stripping feature inside the radius of corotation due to the recent interactions. The interactions offset a fraction of the dust and HI from the main spiral shocks toward larger radii and lead to deeper bending in the iso-velocity contours as shown in Figure~\ref{fig:HI_vLoS}b. In Figure~\ref{fig:HI_vLoS}b, the observed HI intensity map is overlaid on the contours of observed line-of-sight velocity \citep{Wal08}. Figure~\ref{fig:HI_vLoS}b looks similar to the Figure~4 of \citet{Vis80a} \citep[see also][]{Rot75a,Rot75b} except that the angular resolution is two times better and the velocity resolution is ten times higher. As a result, the kinks, as indicators of rapid change in velocity fields, seen in the iso-velocity contours are sharper than that presented in \citet{Vis80a}. 

The general descriptions for the data of four decades old \citep{Rot75a} can be still applied to the new data of much better quality \citep{Wal08}. Theoretical model predicts that kinks in consecutive contours as an indication of shocks fronts, where a rapid change in velocity component perpendicular to shock fronts is expected. It is also expected that density ridges coincide with the turning points in the iso-velocity contours. It has been noted by several authors \citep{Adl96,Vis80a} that the observed turning points along the southwest lie well outside the major density ridge, inconsistent with the expectation of the density wave theory. However, upon closer inspection for Figure~\ref{fig:HI_vLoS}b, one can easily correlate those turning points (as indicated by the small white arrows) in iso-velocity contours with a weak HI ridge. This indicates that those turning points is better associated with the branch-like substructure rather than the main spiral shock. If the HI, which resides in the weak ridge, is tidally stripped from the main shocks toward larger radii, the rotation speed of the gas will be further slowed down due to the conservation of angular momentum and therefore shifts the turning points in iso-velocity contours further out.

Although the shifted turning points in iso-velocity contours shown for the southwest arm are better correlated with the weak HI ridge, Figure~\ref{fig:HI_vLoS}c shows that no apparent inconsistency exists between theoretical prediction and observation. In Figure~\ref{fig:HI_vLoS}c, the observed HI intensity map is superimposed on the line-of-sight velocity obtained from the numerical model $F^{g}_{1.15}c_7$ at $t\approx 1.2$~Gyr. The kinks and the turning points of the calculated iso-velocity contours are well correlated with both the observed northeast and southwest HI major ridges until the corotation radius. As shown in Figure~\ref{fig:HI_vLoS}d, the calculated iso-velocity contours also show good match with the observation except that the bendings of kinks simulated for the southwest arm are not as deep as the observed ones. In fact, one may also find double kinks for the observed iso-velocity contours shown in the northwest quadrant, one is correlated with the major HI ridge and the other one with the weak HI ridge. Beyond the radius of corotation, discrepancy between the simulated and the observed iso-velocity contours becomes apparent. Without external disturbances, the simulated iso-velocity contours tend to converge to the undisturbed circular motions in the region outside corotation radius, while the observed ones do not. Reading directly from the interval of contours, the difference can be up to nearly $25$~km~s$^{-1}$. This again exposes that our simple isolated models for the gasdynamics, especially for the outer disk, is not complete.  The fact that the weak HI ridge can still been seen and has observable kinematic imprint indicates that interaction should be very recent. It is not clear whether the interactions happened roughly 250 Myr ago as suggested by \citet{Yun99} is responsible.

\subsection{The Inner Disk} 
In Figure~\ref{fig:8um_observation}a, the presence of well organized spiral shocks in the inner disk requires an explanation. Since the inner spirals appear to be smoother than the outer spirals, on closer inspection,  one may easily recognize up to four spiral segments in the inner disk. \hhwang{The inner spirals span from $r\approx 3.5$~kpc toward the galactic center and become tightly wound at $r\approx 1.5$~kpc. In Figure~\ref{fig:8um_comparison}, compared to the model with $c_s=7$~km~s$^{-1}$, the inner spiral structures are well developed in the model with $c_s=19$~km~s$^{-1}$ and morphologically consistent with the observation. Furthermore, assuming that those inner dust spirals are due to shock compression at galactic scale, the radially averaged density enhancement (defined as the ratio of the peak surface density to the azimuthally averaged surface density) for the model $c_s=7$~km~s$^{-1}$ is only 1.6, which is significantly lower than that observed, i.e., 2.2. On the other hand, the radially averaged density enhancement are 2.0 and 2.4 for the models with $c_s=11$ and 19~km~s$^{-1}$, respectively.  Since we also expect the observed peak density might be underestimated due to beam smearing (2\arcsec, corresponding to 35~pc at the distance of M81), the model with $c_s=19$~km~s$^{-1}$ is preferred for the origin of inner spirals than the others.} The azimuthal cuts at $r=2.5$~kpc  for these models (at the same temporal point $t\approx 1.2$~Gyr) are shown in Figure~\ref{fig:density_cut_r5p5}d. While only smooth bumps appear near the density peak of stellar spirals in the model $F^{g}_{1.15}c_7$, gas experiences significant compression in the models with higher sound speeds, especially for $F^{g}_{1.15}c_{19}$. Except for the model $F^{g}_{1.15}c_7$, the rapid changes of streaming motions shown on the right are also correlated well with the density peaks in those higher sound speed models, suggesting the presence of shocks. 

The gas situated in the inner disk has a reaction to the imposed spirals in the opposite sense as shown for the outer disk. This phenomenon is best understood in terms of the shifted 4:1 resonance other than the gas response to the external forcing. As shown in Figure~\ref{fig:ultraharmonics}, the 4:1 resonances are \hhwang{significantly} shifted inward in those models with higher sound speeds. The overall spiral structures in gas shrink toward the galactic center with increasing sound speeds. However, the external forcing may modulate the strength of peaks so that the one in the upstream has slightly higher compression than the downstream one. \hhwang{Since the inner stellar spirals are quite open, we suspect that the resonant features dominate over the imposed spirals in the inner disk is simply because the arm strength is not sufficiently strong to allow a transonic point to occur in the path of gas motion. Thus, the small pitch angle assumption is no longer adequate and no shock solution as pursued by \citet{Rob69} can be found. }

%The reason that resonant features dominate over the imposed spirals in the inner disk is simply because the relative arm strength (see Figure~\ref{fig:spiral_amplitude}b) is small inside the bulge. 

The white-dashed ellipses in Figure~\ref{fig:8um_ring_location} mark the radii of ring-like structure for those models as labelled at $t\approx 2.4$~Gyr, which corresponds to 10 orbital times. It shows that the location of ring-like structure has strong dependence on the inner most 4:1 resonances. The ring-like structures located at $r=1$~kpc, as also seen in the inner disk of Figure~\ref{fig:8um_observation}a are in fact composed of tightly wound spirals that are part of 4:1 resonant waves. 

\label{sec:results}
\section{Summary and Discussions} 
\subsection{Discussions}
\hhwang{Figure~\ref{fig:8um_comparison} shows that the region influenced by the spiral potential visually shrinks with the sound speed. For the outer disk, the region where spiral shocks can exist is limited by the spatial extent of the subsonic region around the radius of corotation. For the inner disk, the inward shifted 4:1 resonances also move the overall inner spirals toward the galactic center. These two pressure effects shrink the overall gaseous structures in response to the underlying stellar spiral potential.}

Figure~\ref{fig:8um_comparison}a shows the quasi-regularly spaced substructures reminiscent of those feathers discussed in \citet{LaV06}. We have also shown that the selfgravity is not essential for generating those feathers seen in our two-dimensional, unmagnetized simulations. Since our simulations do not involve the complexities associated with magnetic fields and stellar feedback, this feathering phenomenon as seen in the simulations may be of the wiggle instability found by \citet{Wad04} and presumably of a pure hydrodynamic origin. Within the framework of \citet{Lee12}, \citet{Kim14} applied the normal mode analysis to study the stability of galactic shocks for pure hydrodynamic cases. They concluded that potential vorticities (PVs) generated at deformed shocks may accumulate when the gas parcels successively pass through galactic shocks. As a result, the run away accumulation of PVs leads to the wiggle instability.

In Figure~\ref{fig:potential_vorticity}, the evolution of total PV is shown for the outer disk, in which the outer spiral shocks and the associated wiggle instability reside. The overall increasing rate is proportional to the strength of shocks, with the \hhwang{steepest} slope for the model $F^g_{1.15}c_7$ and the shallowest for $F^g_{1.15}c_{19}$. A closer inspection on the blue curve for $F^g_{1.15}c_7$, one may recognize a rapid increase in the total PV first occurs at roughly 1.7 orbital times, which corresponds to the onset of wiggle instability. The rapid increase stops at 2.4 orbital times, ensuing by a rapid decrease. The oscillation on the top of the overall increasing trend of total PV occurs repeatedly until the end of simulation. The rapid increase is mainly due to the run away deformation of shock fronts, corresponding to the growing modes. The distortion of shocks eventually leads to fragmentation of spirals and therefore run away process stops. Subsequently, the process of reorganizing spiral shocks decreases the total PV due to shock compression and shear reversal \citep{Kim14}. After the shock is well organized by the underlying stellar density waves, the run away process starts over again.

Although the work by \citet{Kim14} provides a physical basis for the wiggles seen in our numerical simulations, linking the wiggle instability to observations for M81 is not straightforward. It has been shown that the two-dimensional wiggle instability is stabilized in three-dimensional simulations due to the flapping motions \citep{Kim06} or the presence of magnetic fields \citep{She06, Kim06}. If magnetic fields are strong in the arm, wiggle instability may be suppressed through reducing the strength of shocks \citep{Lee12, Lee14}. On the other hand, mild magnetic tension may oppose 
Coriolis forces, which would otherwise tend to prevent the coalescence of density clumps along spiral arms, leading to the so-called Magneto-Jeans instability \citep{Kim02, Kim06, She06}. Thus, the balance between selfgravity, epicyclic motions and magnetic tension is delicate for the growth of the feathering instability.  Magnetic fields have been measured for M81 in radio frequencies \citep{Kra89}. Given all other parameters are constrained with observations, it is desirable to investigate further whether or not the presence of the observed strength of magnetic fields would trigger feathering instability as suggested by the normal mode analysis done by \citet{Lee12}. Hence, we find that the physical 
nature of the feathering substructures in M81 is inconclusive. 

In this work, we attribute the inner spiral shocks to the inward shifted 4:1 resonances associated with the presence of WNM. Another possibility that would also shift the resonant features inward is that the pattern speed of the inner spirals is higher than the value adopted in our models. Thus, observational determination of pattern speed for the inner arms would be crucial to our understanding of the origin of spiral structure in M81.

The complicated HI tidal tails observed in the outskirts of M81 show clear signature of interactions \citep{Yun94}, which are not considered in this work. Simulations suggest that M81 may have interacted  with M82 and NGC~3077 roughly 200 to 400 Myrs ago \citep{Tho93,Yun99}. The close encounters may disturb the disk structure and the gas distribution. Therefore, we do not expect that our isolated models for M81 can reproduce all the observed features. Instead, we consider the deviations from the our bi-symmetric models as a result of interactions or physics missing in the models. We identify that the ultra-high streaming motions observed in the northeast arm are not consistent with the observed arm strength. Especially, we associate the outward shifted turning points with the weak HI density ridge (see Figure~\ref{fig:HI_vLoS}b) and argue that there is no apparent contradiction to the prediction of the theory of spiral density waves based on the good fit on the locations of outward bending kinks seen in iso-velocity contours. Since this kinematics feature has no counterpart in the northeast arm, the weak HI density ridge is interpreted as a tidally stripped feature, perhaps due to the interaction with NGC~3077 according to the simulated orbits suggested by \citet{Yun99}.  Numerical simulations involving both live stellar and gaseous disks would be desirable to justify that close encounter with companions could cause the observed deviation from the results obtained from isolated models. 

\label{sec:summary_discussion}

\subsection{Summary}
Hydrodynamic simulations with a parameter space constrained by observations are performed to study the gas response to the stellar spiral density waves in M81. Based on the halo-bulge-disk model constructed for M81, an unstable spiral mode, which is consistent with the stellar disk, is found to fit with observations through the modal analysis reported in \citetalias{Fen14}. Therefore, the spiral mode together with its associated pattern speed and its modulation of arm strength derived in \citetalias{Fen14} are taken as the inputs in this work. 

With the QSSS hypothesis, the imposed spiral density waves rotate about the galactic center with a constant pattern speed, $\Omega_p=25.5$~km~s$^{-1}$~kpc$^{-1}$ \citepalias{Fen14}. Within the uncertainty of observations, the gas response to three different arm strengths is explored, aiming to further narrow the range of possible arm strengths. The effective sound 
speeds, which are taken to reflect the turbulent nature of the interstellar medium, are constrained by the values obtained by two-Gaussian decomposition for the stacked HI line profile. The result obtained for M81 suggests the existence of CNM with  $c_s=7$~km~s$^{-1}$ and WNM with $c_s=19 $~km~s$^{-1}$ \citep{Ian12}. These two values together with the value, $c_s=11$~km~s$^{-1}$, estimated from the single-Gaussian fit are chosen to model the effective isothermal gases residing in the disk. By simulating the CNM and the WNM separately, we implicitly assume the responses of gases with different sound speeds are dynamically decoupled, while those models with $c_s=11$~km~s$^{-1}$ are representative when the dynamics of CNM and WNM is well coupled. The mass distribution used to initialize the gas disk is also derived from the recent HI observation \citep{Wal08}. Given these observational constraints, a good fit of the gas response to the underlying spiral density waves with observations may help identify possible mechanisms and conditions that lead to the observed structures in gas, while a poor fit may suggest that the physics involved is incomplete. In total, 18 simulations are performed to explore the parameter space comprising three different arm strengths, three different sound speeds, and with or without the inclusion of selfgravity. The major conclusions are summarized as follows. 

\begin{figure}
\epsscale{1}
\plotone{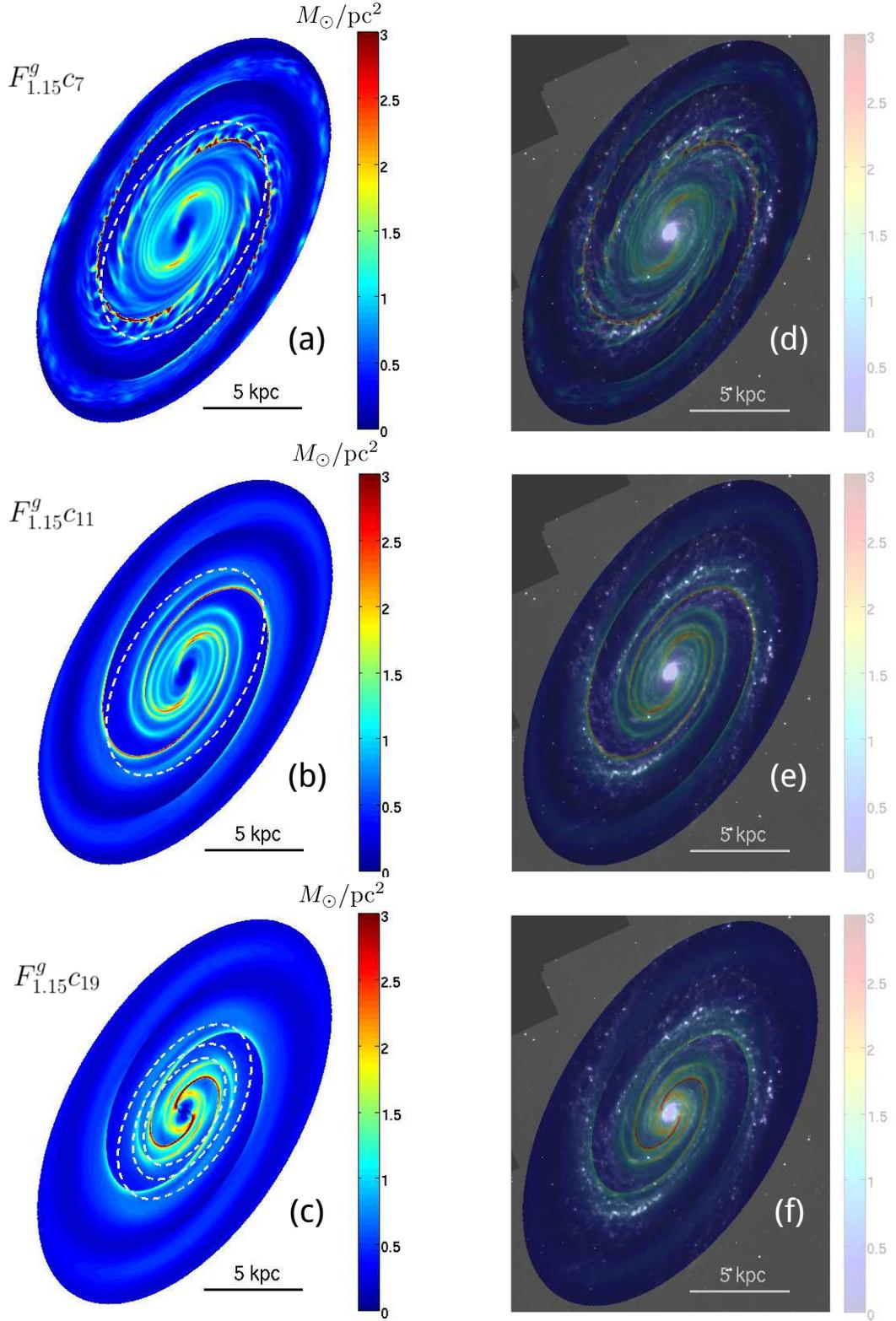}
\caption{{\it Left column:} The snapshots of the simulated dust emission at $t\approx 1.2$~Gyr, corresponding to 5 revolution times of the spiral pattern. The corresponding model of each image is labelled in the upper-left corner. The white-dashed ellipses mark the radii of inner 4:1 resonances in the plane of the galaxy (see the discussion in Section~3.2). {\it Right column:} Direct comparisons by overlaying the simulation results shown in the left column over the 8~$\mu$m image shown in Figure~\ref{fig:8um_observation}a \label{fig:8um_comparison}}
\end{figure}

\begin{figure}
\epsscale{1}
\plotone{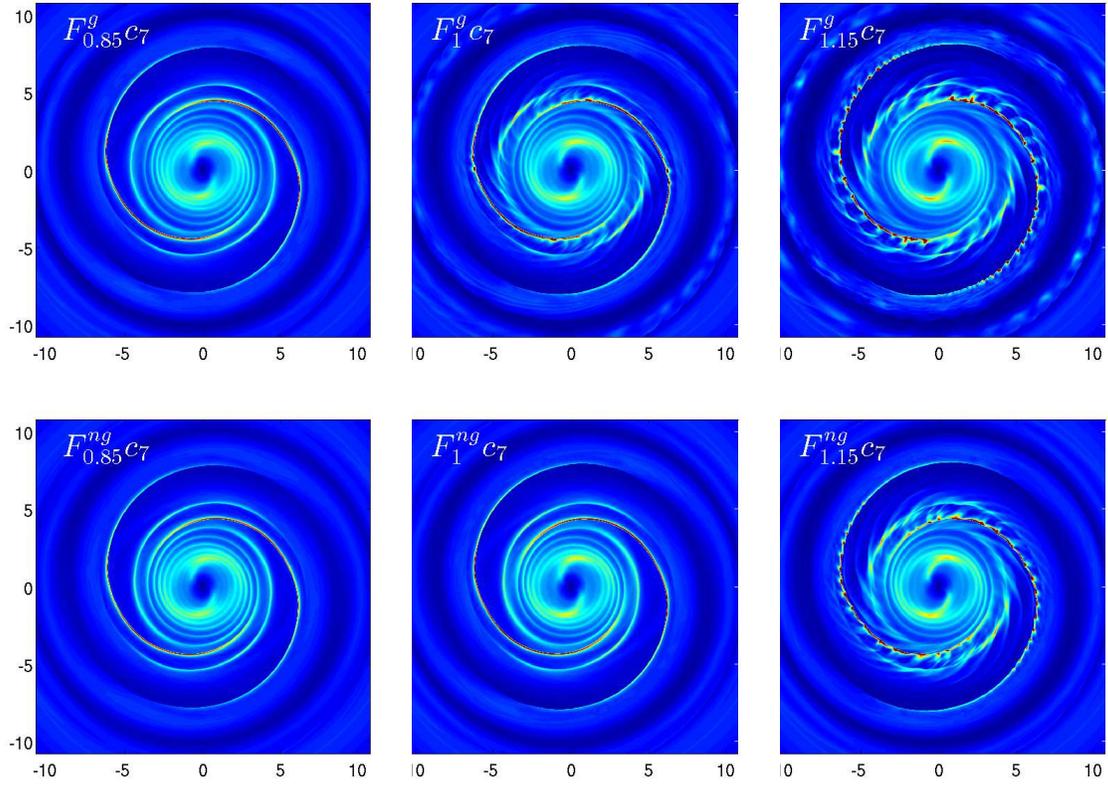}
\caption{Snapshots of surface density at $t \approx 1.2$~Gyr for selfgravitating (upper panel) and non-selfgravitating (lower panel) models as labelled in the upper-left corner of each image. The units of axes are in kpc and the color code used is the same as that used for the left column of Figure~\ref{fig:8um_comparison}. \label{fig:image_array_c7}}
\end{figure}

\begin{figure}
\epsscale{0.7}
\plotone{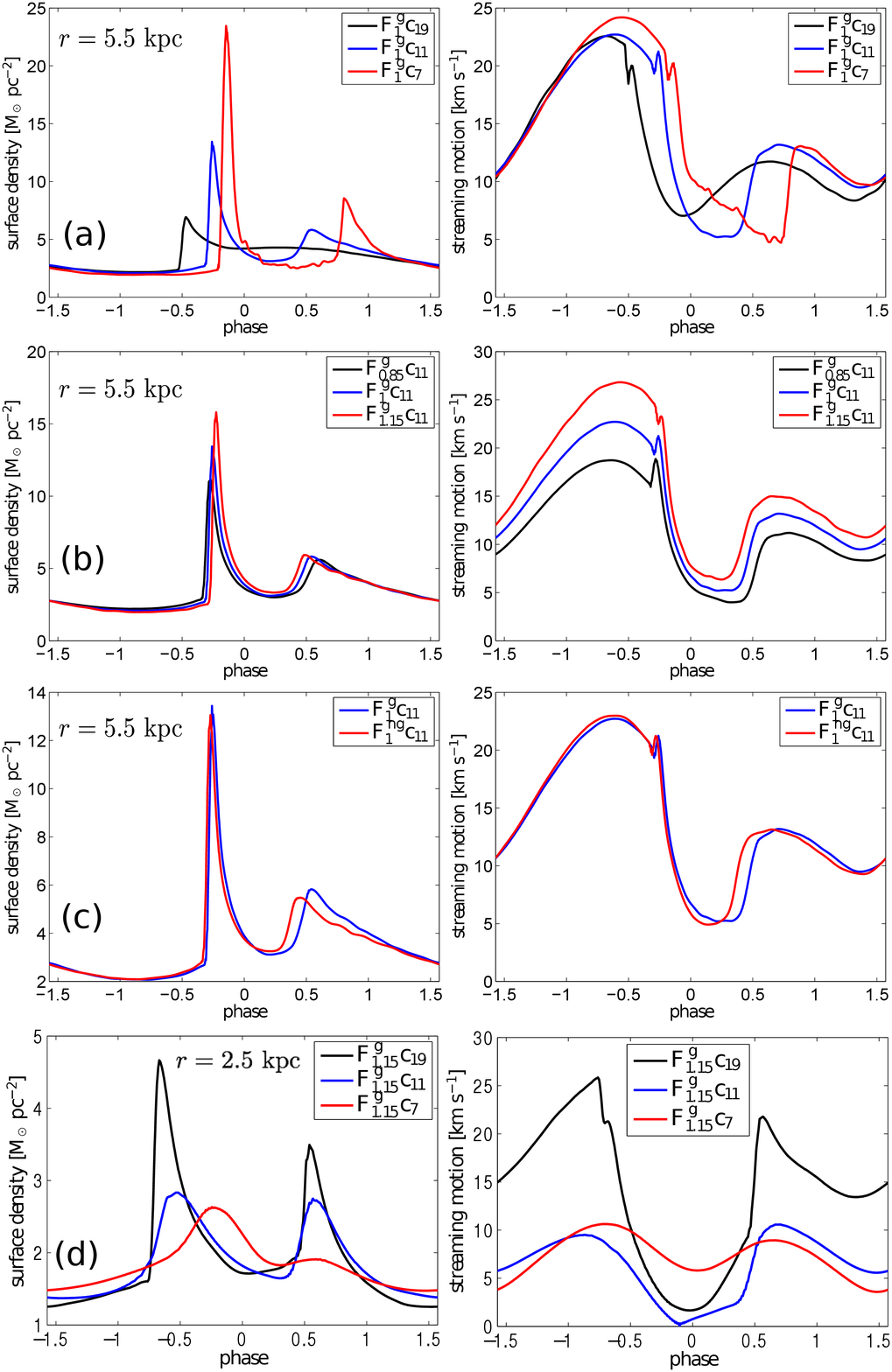}
\caption{{\it left column:} Azimuthal cut of surface density at the radii $r=5.5, 2.5$~kpc at the temporal points $t=450$~Myr for (a), (b), (c) and $t=1.2$~Gyr for (d). The zero phase corresponds to the peak surface density of the imposed spirals at the radius. To avoid duplicated information, only the range $[-\pi/2~\pi/2]$ is shown. {\it right column:} The corresponding streaming motions, measuring the magnitude of velocity departure from the basic state. \label{fig:density_cut_r5p5}}
\end{figure}

\begin{figure}
\epsscale{1}
\plotone{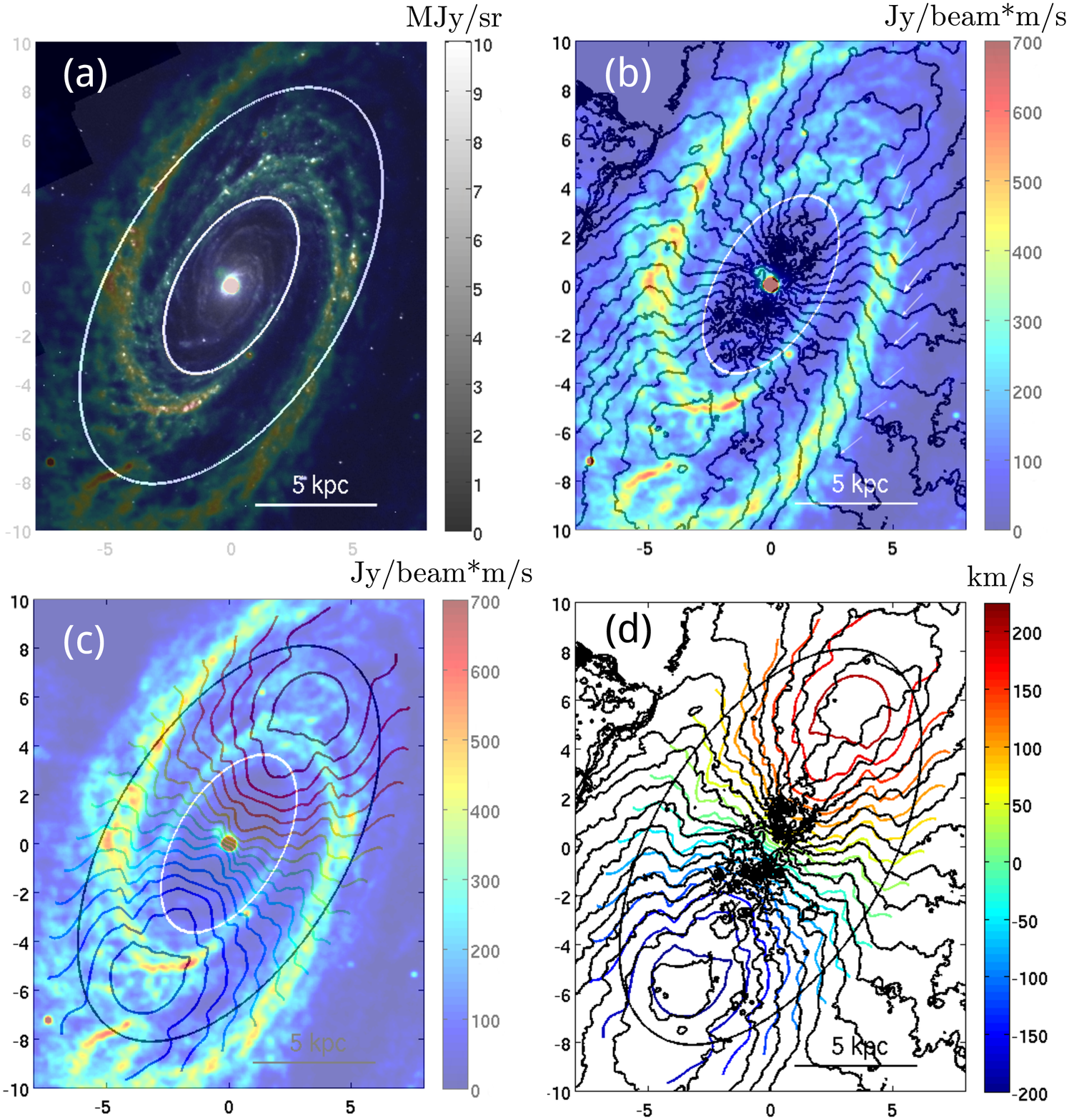}
\caption{(a) Superposition of the 8~$\mu$m \citep{Wil04} and 21~cm \citep{Wal08} images. The white ellipses with increasing radius correspond to the radii of 4~kpc and corotation (9.08~kpc), respectively. (b) Superposition of the observed maps of HI intensity (color coded) and iso-velocity contours \citep{Wal08}. A system velocity -39.4~km~s$^{-1}$ is subtracted from the first moment map. The contours are displayed every 25~km~s$^{-1}$ for the velocity range -200 to 200~km~s$^{-1}$. The southern loop seen along the major axis corresponds to -200~km~s$^{-1}$ and the northern one to 200~km~s$^{-1}$. The white ellipse marks the radius of 4~kpc. The small white arrows mark the locations of turning points in the iso-velocity contours. (c) Superposition of the observed HI intensity map and the simulated iso-velocity contours for the best fit model $F^{g}_{1.15}c_7$ at $t=1.2$~Gyr. The simulated velocity field is smoothed using a beam size (12\arcsec) comparable to the observation. The map of iso-velocity contour is calculated using an inclination angle $58^{\circ}$. The levels of contours is the same as those adopted in (b). The black and white ellipses mark the radius of corotation and a 
radius of 4~kpc respectively. (d) Superposition of the observed and the simulated ($t=1.2$~Gyr) iso-velocity contours. \label{fig:HI_vLoS}}
\end{figure}

\begin{figure}
\epsscale{0.5}
\plotone{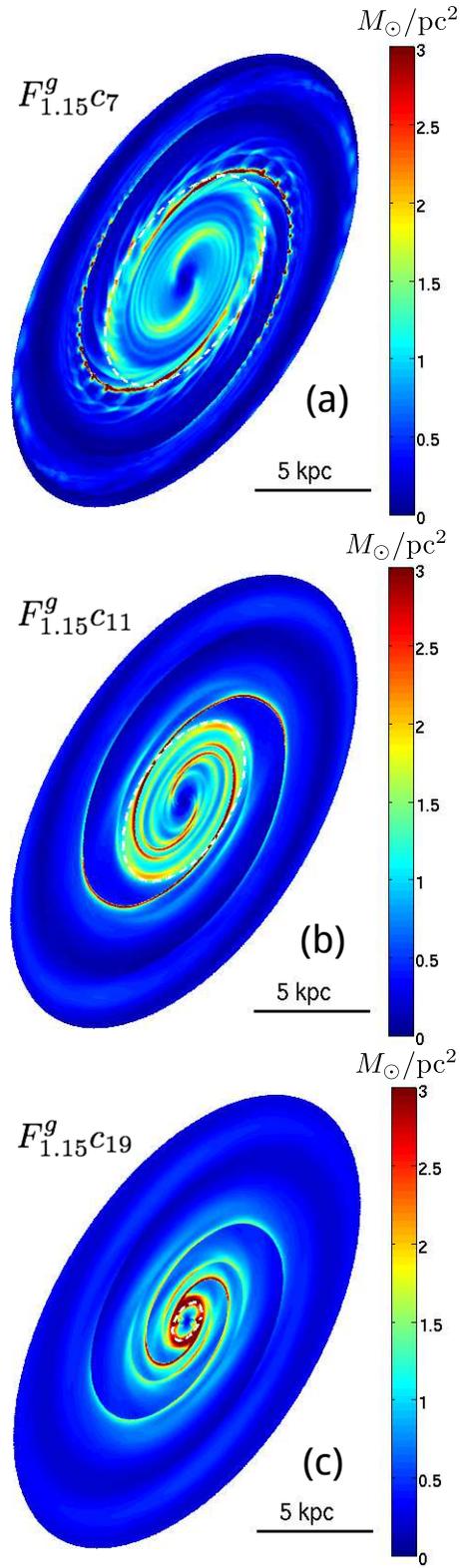}
\caption{ The snapshots of the simulated dust emission at $t\approx 2.4$~Gyr, corresponding to 10 revolution times of the spiral pattern. The model corresponding to each image is denoted in the upper-left corner. From top to bottom,  the white-dashed ellipses mark the radii 5, 3.8 and 0.9~kpc in the plane of galaxy, respectively. \label{fig:8um_ring_location}}
\end{figure}

\begin{figure}
\epsscale{0.8}
\plotone{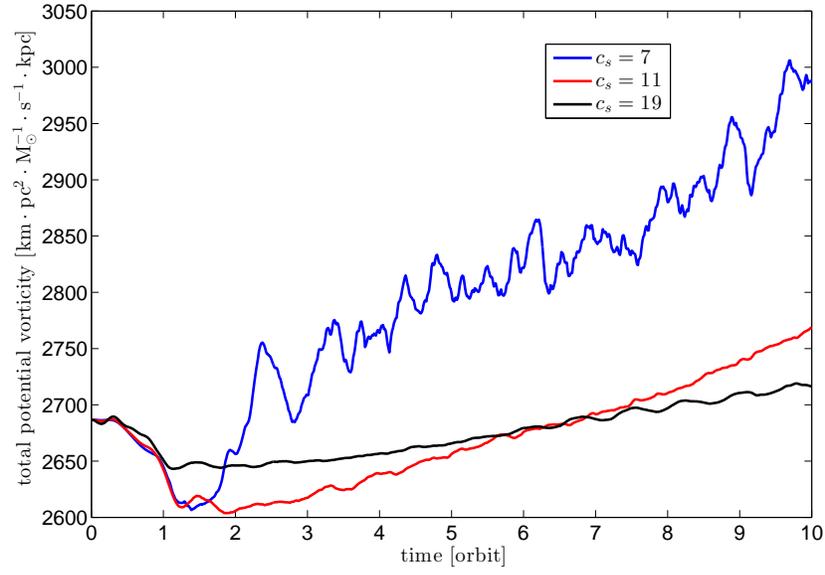}
\caption{Evolution of total potential vorticities integrated over the regions defined by $4<r<10.8$ for the models $F^{g}_{1.15}c_7$ (blue), $F^{g}_{1.15}c_{11}$ (red) and $F^{g}_{1.15}c_{19}$ (black).  \label{fig:potential_vorticity}}
\end{figure}

\begin{enumerate}
\item In both the 8~$\mu$m and 21~cm images, the ring-like structure situated just outside the radius $r=4$~kpc is interpreted as a feature of the 4:1 resonance associated with the CNM. The location of the simulated ring structures shrinks toward the galactic center with increasing sound speeds due to the inward shifted 4:1 resonances. 

\item The spiral shocks shown in the inner disk are identified as resonant features rather than the type of gas response to the spiral forcing pursued by \citet{Rob69}. The 4:1 resonances associated with the WNM explain the origin of the four-armed spirals residing in the inner disk as well as the tightly wound spiral structure observed at $r\approx 1.5$~kpc. 

\item Within the uncertainty of the observed arm strength, the level of simulated streaming motions is consistent with that observed for the southwest arm and favors those models with stronger arm strength, i.e., $x=1, 1.15$.

%comparison between the simulated streaming motions and the observed values prefers those models with higher arm strengths, i.e., $x=1, 1.15$. }

\item Detailed comparisons between the data observed at wavelengths of 8~$\mu$m and 21~cm together with those obtained from simulations enable us to identify possible signatures of previous interactions in the near past. Although the model $F^{g}_{1.15}c_7$ shows good fit to the southwest spiral of dust (Figure~\ref{fig:8um_comparison}d), the corresponding HI tail stretches further out beyond the corotation radius (Figure~\ref{fig:HI_vLoS}a). The lack of counterpart emission in 8~$\mu$m for the neutral gas situated beyond the corotation radius reflects that the origin of this HI tail might be external to the galactic disk. Moreover, those turning points shown in the iso-velocity contours are found to be correlated well with the weak HI density ridge in the new data (Figure~\ref{fig:HI_vLoS}b), suggesting that the unique branch-like structure seen downstream the southwest arm might be the material tidally stripped away from the main spiral. Therefore, the shifted turning points from the major southwest spiral are considered to be a natural consequence of  interactions. A signature of interactions can be also found in the northeast arm since simulation results show that the anomalously high streaming motions observed in this arm cannot be accounted by the observed arm strength. Interaction scenarios might be a natural direction to explain the asymmetry between the two arms. 

\item Gases of different sound speeds respond differently to the spiral density waves in M81. Simulations show that spatial separation between the density peaks of CNM ($c_s=7$~km~s$^{-1}$) and WNM ($c_s=19$~km~s$^{-1}$) can be more than 2~kpc in the outer disk and may be detectable using a stacking technique for the HI data. 

\item Our simulation results suggest that the selfgravity of gas does not play an important role in triggering wiggle instability. The nature of this type of instability may be of pure hydrodynamics as suggested by \citet{Kim14}. However, the nature of the feathers observed in M81 is uncertain based on this work. A further study that takes into account the effects of magnetic fields and the vertical structure of gas disk is required to distinguishing between different 
theoretical possibilities.

\item Direct confrontation between numerical models and the observations obtained from different wavelengths have shown that simulations with single isothermal sound speed alone cannot explain all the structures seen in the gas and dust. This 
provides further support for the coexistence of CNM and WNM in M81. However, the exact distribution of the neutral gases of different velocity dispersions in M81 is not clear, and their determination would be desirable in future 
observational studies. The spiral structures in M81 are not perfectly bi-symmetric with respect to the galactic center. Therefore, we do not expect the results obtained from simulations such as those carried out in this paper 
would adequately fit both the inner and outer structure within the same theoretical framework.  Although tidal 
interactions with neighboring companion galaxies are likely to play a role in forming the structure in the outer disk 
of M81, it is unclear whether their inclusion alone would be sufficient.  The degree to which galaxy interactions or 
physics beyond those considered in this paper should be considered in future studies. 
\end{enumerate}

 \acknowledgments
 \hhwang{The authors would like to acknowledge the support of the Theoretical Institute for Advanced Research in Astrophysics (TIARA) based in Academia Sinica Institute of Astronomy and Astrophysics (ASIAA). HHW is grateful to Prof. Woong-Tae Kim for the valuable inputs and discussions during the EANAM meeting. The authors thank the referee for comments which helped to improve the clarity and presentation of this paper. Thanks to Mr. Sam Tseng for assistance on the computational facilities and resources (TIARA cluster).} 
 \bibliography{M81_hydro_bib}
\clearpage

%\begin{figure}
%\epsscale{0.6}
%\plotone{velocity_diff.eps}
%\caption{Relative circular velocity with respect to the spiral pattern as a function of galactocentric radius. \label{fig:velocity_diff}}
%\end{figure}

\end{document}